\documentclass[amsfonts,showpacs,tightenlines,aps,12pt,floatfix]{revtex4}
\usepackage{bm}
\usepackage{epsfig}
\begin{document}


\title{Consequences of the BaBar $e^+e^-\to\pi^+\pi^-$ Measurement 
for the Determination of Model-Dependent $\rho$-$\omega$ Mixing Effects
in $\Pi_{\rho\omega}(m_{\rho}^2)$ and $(g-2)_{\mu}$}.
\author{Carl E. Wolfe}
\email[]{wolfe@yorku.ca}
\affiliation{Department of Physics and Astronomy, York University,
4700 Keele St., Toronto, ON CANADA M3J 1P3}
\author{Kim Maltman}
\email[]{kmaltman@yorku.ca}
\affiliation{Department of Mathematics and Statistics, York University,
4700 Keele St., Toronto, ON CANADA M3J 1P3}
\altaffiliation{CSSM, Univ. of Adelaide, Adelaide, SA 5005 AUSTRALIA}
\date{\today}

\begin{abstract}
We update our analysis of $\rho$-$\omega$ mixing effects
in the pion 
form factor to incorporate the recently published BaBar $e^+e^-\to\pi^+\pi^-$
cross-sections.  
The implications for $\tau$-decay-based Standard Model estimates of the leading 
order hadronic contribution, $[a_\mu]_{had}^{LO}$, to the anomalous magnetic 
moment of the muon, and for the extraction of the off-diagonal vector meson 
self-energy matrix element, $\Pi_{\rho\omega}(m_\rho^2)$, are discussed.
\end{abstract}


\maketitle

In the following we update the analysis performed in Ref.~\cite{wm10}
of the isospin-breaking (IB)
$\rho-\omega$ mixing correction required in order to use $\tau$-decay-based
data instead of electroproduction data in the evaluation
of $[a_\mu]^{LO}_{\rm had}$, the leading order hadronic vacuum 
polarization contribution to the anomalous magnetic moment 
of the muon, $a_\mu\equiv (g-2)_\mu/2$.  This update focusses on the BaBar
electroproduction data~\cite{babarpipi09, babar09epaps} since it was not
released until shortly after publication of the previous analysis.  

As is well known, 
several recent measurements of the $e^+e^-\to\pi^+\pi^-$ cross-section
\cite{babarpipi09, cmd203, kloe04, kloe08, snd05pipi, snd06pipi, cmd207} 
together yield estimates of $[a_\mu]^{LO}_{\rm had}$ which are 
consistent with one another, but lead to Standard Model (SM)
predictions for $a_\mu$ deviating from the 
BNL E821 experimental result \cite{bnlgminus2}
by $\sim 3.2-3.6\sigma$~\cite{davier2007,passera07,hmnt2007,hmnt2010,jegerlehner09,davieretal09,davieretal10,hoecker10}. 
In contrast, evaluating $[a_\mu]^{LO}_{\rm had}$
using $\tau$ decay data in place of isovector electroproduction 
data~\cite{aleph97, opal99, cleo00, aleph05, dhz06, belle08taupipi} 
yields a SM prediction for $a_\mu$ differing from experiment by 
only $\sim 1.9-2.4\sigma$~\cite{davier2007, passera07, jegerlehner09,davieretal09,davieretal10,hoecker10}.
Use of the $\tau$ decay data requires that a number of small 
IB corrections to the CVC relation be taken into account.  
These corrections have been extensively studied in 
Refs.~\cite{cen01, cen02, gj03, fflt06, fflt07,davieretal09} and
are believed to be well understood.  We denote these corrections
collectively by $[\delta a_\mu]^{LO}_{\rm had}$ and focus in 
what follows on the particular contribution arising from $\rho-\omega$ mixing, 
$[\delta a_\mu]^{LO}_{\rm had; mix}$, which
is defined explicitly in Ref.~\cite{wm10}.  

An important observation made in Ref.~\cite{mw06} was that the generic structure
of the $\rho-\omega$ interference contribution to $F_{\pi}(s)$ 
introduces strong fit-parameter-sensitive cancellations, 
and hence significant model dependence, into the integral corresponding to 
$[\delta a_{\mu}]_{had;mix}^{LO}$.
Our analysis thus employs a range of models for $F_\pi(s)$, all
having some basis in phenomenology. These are
the Kuhn-Santamaria (KS) model~\cite{ks90}, 
the Hidden Local Symmetry (HLS) model~\cite{hlsbasic, hlsgood}, 
the Gounaris-Sakurai (GS) model~\cite{gs68}, and a modified version of the 
GP/CEN model~\cite{gp97, cen02}. (Detailed descriptions of 
the models can be found in section II of Ref.~\cite{wm10}.)
Refs.~\cite{mw06, wm10} show that it is necessary to consider 
such a range of models if one wishes to properly assess the model dependence of 
$[\delta a_{\mu}]_{had;mix}^{LO}$, and, from this, 
the uncertainty in the $\pi\pi$ contribution to 
the $\tau$ decay-based estimates of $[a_\mu]^{LO}_{\rm had}$.  

Shortly after the publication of Ref.~\cite{wm10} 
the BaBar collaboration released the data corresponding to its measurement 
of the 
$e^+e^-\to\pi^+\pi^-(\gamma)$ cross-section, using the initial-state 
radiation method, from threshold to 3 GeV \cite{babarpipi09,babar09epaps}.  
Compared to the electroproduction data sets described and used in 
Ref.~\cite{wm10} (CMD-2~\cite{cmd203, cmd207}, SND~\cite{snd05pipi,snd06pipi}, 
and KLOE~\cite{kloe04,kloe08})
the BaBar data offers considerably increased statistics, including 15 data 
points in the interference region (770-800 MeV), as well as generally lower
statistical and systematic errors.
The BaBar data distinguishes itself from its predecessors, however, in that
the value of $a_\mu$ computed using it as the source of the 
$\pi\pi$ contribution to $[a_\mu]^{LO}_{\rm had}$
more closely corresponds to the experimental and $\tau$ decay
based values, deviating from the experimental value by only 
$2.4\sigma$~\cite{davierbabar}.  

As before, we perform fits to the BaBar data set using the models indicated
above.  Although the BaBar data extends up to 3 GeV, 
only the low-energy part of this data is relevant to analyzing $\rho-\omega$
mixing.  We, therefore, limit our analysis to the
maximum $e^+e^-$ center-of-mass energy of 970 MeV 
employed in our previous analysis.
The results quoted below for 
$[\delta a_\mu]^{LO}_{had;mix}$ are insensitive to modest changes in this
choice of endpoint.
All results correspond to the bare
form factor (i.e. with the effects of vacuum polarization removed).
Details of the fit procedure, including all input values, are unchanged from
Ref.~\cite{wm10}.
Fit results for each model are shown in Table \ref{table1}.  The fit 
parameters are the $\rho$ mass and width, $m_\rho$ and $\Gamma_\rho$, the 
complex coefficient of the $\omega$ contribution, $\delta$, the coefficient
of the $\rho^\prime$ term, $\beta$, and the HLS model parameter, $a_{HLS}$.
A blank entry indicates that a fit parameter is inapplicable
to that particular model.  For the GP/CEN$^+$ and GP/CEN$^{++}$
models, the effective value of $\Gamma_{\rho}$ is shown in brackets to 
highlight that it is in fact $\delta\Gamma_{\rho}$, an offset from the nominal
chiral effective theory $\rho$ width, which is the actual fit parameter.  

\begin{table}[h]
\caption{\label{table1}
Results of fits to the BaBar 2009 data.}
\vskip .1in
\begin{tabular}{|c||c|c|c|c|c|} \hline Parameter & KS & HLS & GS & GP/CEN$^+$ & GP/CEN$^{++}$ \\
\hline
$m_{\rho}$ (MeV) & 772.11$\pm$0.30 & 773.48$\pm$0.29 & 774.29$\pm$0.30 & 775.87$
\pm$0.29 & 775.87$\pm$0.29 \\
$\Gamma_{\rho}$ (MeV) & 147.56$\pm$0.54 & 149.68$\pm$0.57 & 149.87$\pm$ 0.57 & (148.66) & (148.65) \\
$\delta\Gamma_{\rho}$ (MeV) & - & - & - & 1.33$\pm$0.46 & 1.32$\pm$ 0.46\\
$|\delta|$ ($10^{-3}$) & 1.89$\pm$0.03 & 1.99$\pm$0.03 & 1.96$\pm$0.03 & 2.30$\pm$0.
03 & 1.98$\pm$ 0.03 \\
${\rm Arg}(\delta)$ (deg) & 9.4 $\pm$ 1.2 & 10.3 $\pm$ 1.1 & 10.3 $\pm$ 1.1 & 10.8 $
\pm$ 1.1 & 10.8 $\pm$ 1.1 \\
$\beta$  & -0.152$\pm$0.002 & - & -0.088$\pm$0.002& - & - \\
$a_{HLS}$ & - & 2.3989$\pm$0.007 & - & - & - \\
\hline
$\chi^2$/dof & 392/250 & 320/250 & 322/250 & 413/251 & 414/251 \\
\hline
\end{tabular}
\end{table}

Comparing the results of Table \ref{table1} with those of Tables I-IV in 
Ref.~\cite{wm10}, we see that the BaBar data yields a $\rho$ width larger by 
1-7 MeV (depending on the specific data set and model)
and a reduced $\rho$-$\omega$ mixing phase.  
Reasonable $\chi^2$/dof results are obtained despite the reduced scale of 
statistical errors in the BaBar data relative to the other data sets.

\begin{table}[h]
\caption{\label{amutable}
$[\delta a_{\mu}]^{LO}_{had;mix}\times 10^{10}$ for the models discussed in the
text and
the CMD-2, SND, and KLOE $e^+e^-\to\pi^+\pi^-$ cross-sections.}
\vskip .1in
\begin{tabular}{|c||c|c|c|c|c|}
\hline
Experiment & KS & HLS & GS & GP/CEN$^+$ & GP/CEN$^{++}$ \\
\hline
CMD-2(94) & $3.8\pm 0.6$ & $4.0\pm 0.6$ & $2.0\pm 0.5$ & $2.0\pm 0.5$ & $1.8\pm
0.4$ \\
CMD-2(98) & $4.0\pm 0.6$ & $4.6\pm 0.6$ & $2.5\pm 0.5$ & $2.2\pm 0.4$ & $2.1\pm
0.4$ \\
SND & $4.2\pm 0.4$ & $4.3\pm 0.4$ & $2.2\pm 0.3$ & $1.9\pm 0.3$ & $1.7\pm 0.3$ \\
KLOE(02) & ($2.2\pm 0.6$) & $4.2\pm 0.7$ & $2.2\pm 0.6$ & ($0.5\pm 0.8$) & ($0.3 \pm
 0.8$) \\
BaBar(09) & $5.0\pm 0.2$ & $5.0\pm 0.2$ & $2.9\pm 0.2$ & $2.6\pm 0.2$ & $2.4 \pm 0.2
$ \\
\hline
\end{tabular}
\end{table}

The values obtained for $[\delta a_{\mu}]^{LO}_{had;mix}$ using the BaBar 
data and for each of the models
considered are shown in Table~\ref{amutable}, along with 
the values from the data sets used in Ref.~\cite{wm10}.  The latter are included
for ease of comparison.  The BaBar data
yields somewhat larger central values, along with reduced 
errors.  However, as before \cite{wm10, mw06}, the variation in the values 
of $[\delta a_{\mu}]_{had;mix}^{LO}$ across the 
various models is greater than the experimental uncertainty produced by 
any single model.  In arriving at a final assessment of our results for
$\left[\delta a_\mu\right]^{LO}_{had;mix}$, we have adopted the
view that, since all the models considered have a
reasonable basis in phenomenology, all results corresponding
to a given data set and given model which produce an
acceptable quality fit are to be included in the assessment.
(Those entries in brackets in Table~\ref{amutable} correspond to poor
quality fits and are not included in our final result.)
We thus first perform a weighted average over all experiments
for each separate model, and then take the average (half the
difference) of the maximum and minimum values allowed by the resulting
error intervals for the different models to define our central values
(model-dependence-induced uncertainties).
The updated combined assessment, now including the BaBar results, is 
\begin{equation}
\label{CVall}
\left[\delta a_{\mu}\right]^{LO}_{had;mix} = (3.5\pm 1.5_{model}\pm 0.2_{data})\times 10^{-10}.
\end{equation}
The central value has increased by $0.4\times 10^{-10}$ and the data 
error has decreased by $0.1\times 10^{-10}$ compared to the value 
reported in \cite{wm10}.

The value shown in Table~\ref{amutable} obtained using BaBar data and
the GS model is compatible with the GS result reported 
in Ref.~\cite{davierbabar}.  The KS model result, however, is not,
the KS and GS results for $\left[\delta a_{\mu}\right]^{LO}_{had;mix}$
differing significantly in Table~\ref{amutable} but being the
same in Ref.~\cite{davierbabar}. The source of this apparent discrepancy 
is that two distinct `KS' models have in fact been employed:
the one we denoted KS above, and the alternate version
used in Ref.~\cite{davierbabar}, which we call KS$^\prime$.  As discussed in 
Ref.~\cite{wm10} these two models differ in the $s$-dependence 
assumed for the $\rho$-$\omega$ mixing contribution to $F_\pi (s)$.  
We have confirmed that the alternate, KS$^\prime$, form indeed yields 
results for $\left[\delta a_{\mu}\right]^{LO}_{had;mix}$ compatible 
with those of the GS model. In fact, it turns out that the presence or absence 
of the extra $s/m_\omega^2$ factor (which is what distinguishes
the KS and KS$^\prime$ model forms) is also the key feature distinguishing
those models which yield `high' values of 
$\left[\delta a_{\mu}\right]^{LO}_{had;mix}$ (KS, HLS) from those which 
yield `low' values (GS, GP/CEN$^{++}$).
The data, in the narrow range of $s$ over which $\rho-\omega$ interference
is significant, is incapable of distinguishing between these differing
$s$-dependences.  While such differences have only a
very small impact on the values of the model fit parameters, the presence or 
absence of the factor of $s/m_\omega^2$ strongly affects the very close 
cancellation occurring in the weighted integral for 
$\left[\delta a_{\mu}\right]^{LO}_{had;mix}$.
Since there is, at present, no compelling 
theoretical argument favouring one choice of $s$-dependence over the other
in the interference region, 
we adopt the view that the unknown $s$-dependence of the mixing term 
must be treated as an additional source of uncertainty for
$\left[\delta a_{\mu}\right]^{LO}_{had;mix}$.  This
uncertainty significantly increases the total error on
$[a_{\mu}]^{LO}_{had; mix}$.
\bigskip

As explained in Refs.~\cite{mow,wm10}, analysis of the
electroproduction data in the interference region 
also allows one to extract
the off-diagonal $\rho-\omega$ element of the vector meson self-energy matrix,
$\Pi_{\rho\omega}(q^2)$, and the isospin-breaking coupling ratio 
$G \equiv g_{\omega^I\pi\pi}/ g_{\rho^I\pi\pi}$, with $g_{\omega^I\pi\pi}$ and 
$g_{\rho^I\pi\pi}$ the isospin-pure $\pi\pi$ couplings of the $\rho$ and 
$\omega$ mesons.  $\Pi_{\rho\omega}(q^2)$ is of interest, for example,
for meson-exchange models of IB in the NN interaction.  The procedure for 
performing this determination has been described in detail in 
Refs.~\cite{mow, wm10}.

\begin{figure}[ht]
\begin{minipage}[b]{0.48\linewidth}
\centering
\includegraphics[scale=0.30, angle=-90]{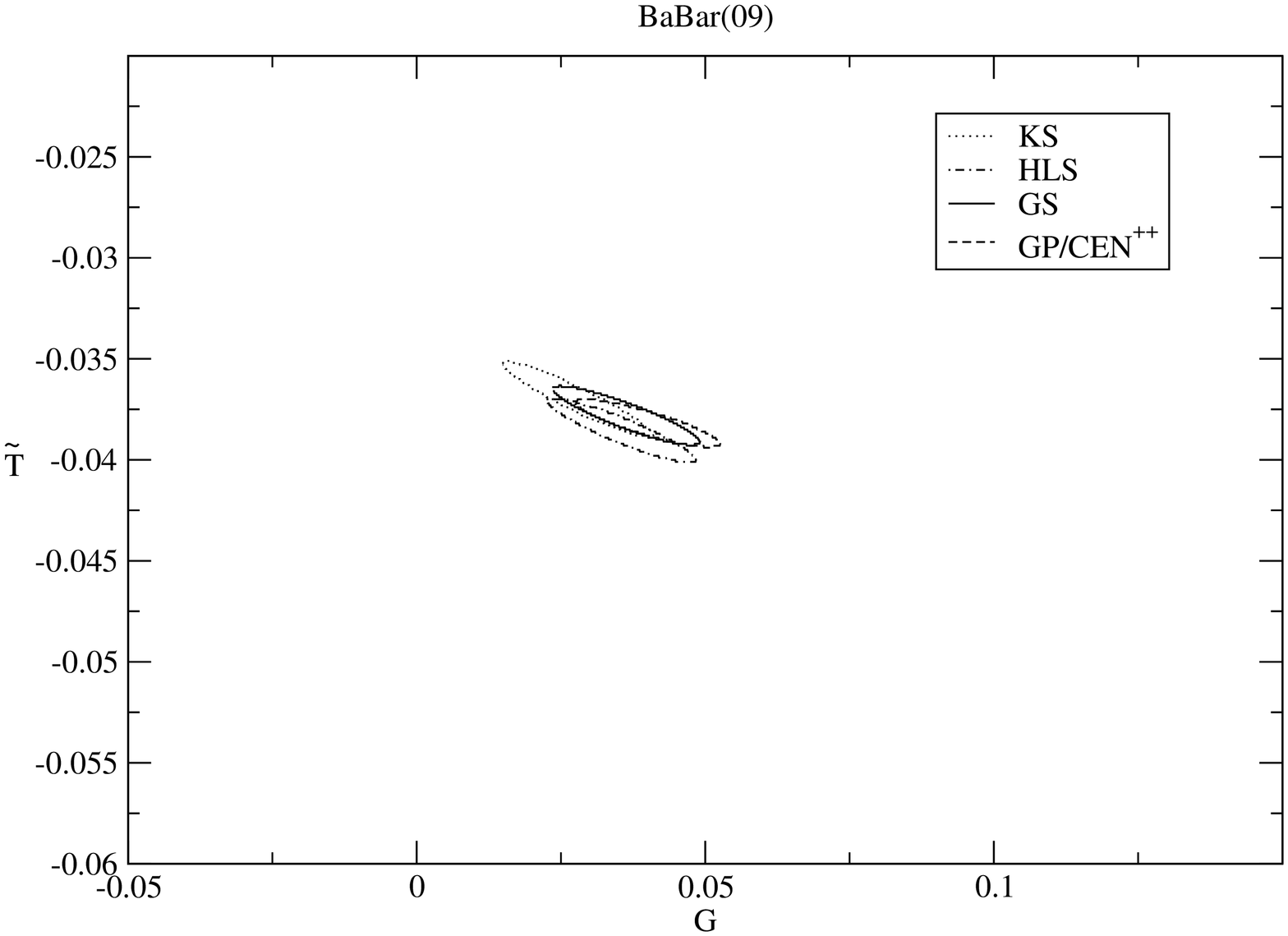}
\caption{BaBar(09) $G$ and $\tilde{T}$ one-sigma regions.}
\label{Babar09MOW}
\end{minipage}%
\hspace{0.5cm}
\begin{minipage}[b]{0.48\linewidth}
\centering
\includegraphics[scale=0.30, angle=-90]{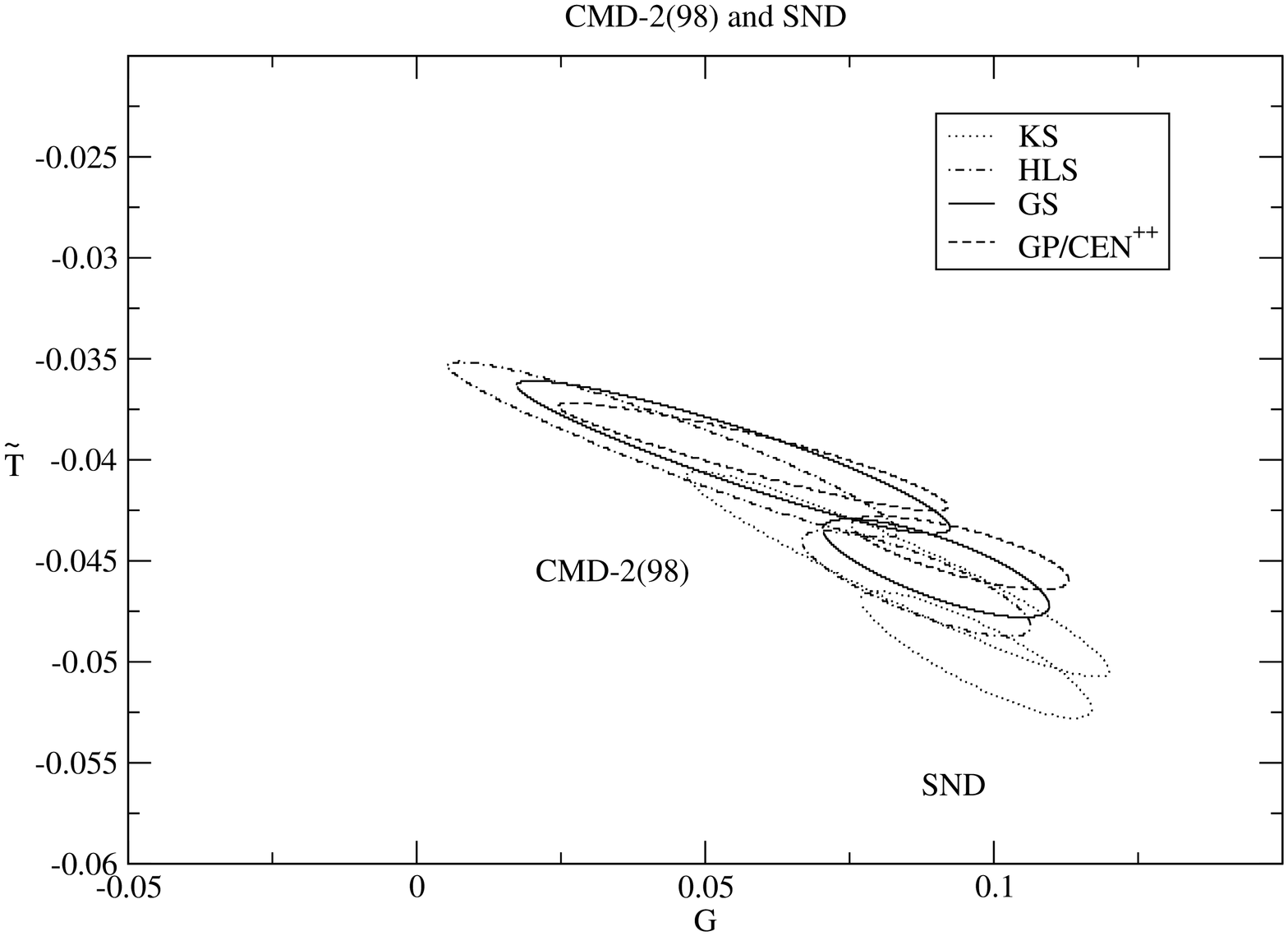}
\caption{CMD-2(98) and SND $G$ and $\tilde{T}$ one-sigma regions.}
\label{CMD2SNDMOW}
\end{minipage}
\end{figure}

The separation of mixing and direct $\omega\rightarrow\pi\pi$
contributions depends on the model used for the broad
$\rho$ contribution to $F_{\pi}(s)$.  We report in 
Table~\ref{table3} the results
for $\phi$ (the Orsay phase), $G$, and 
$\tilde{T} \equiv \tilde{\Pi}_{\rho\omega}(m_\rho^2)/\hat{m}_\rho\Gamma_\rho$ 
(with $\tilde{\Pi}_{\rho\omega}$ 
the real part of $\Pi_{\rho\omega}$ and $\hat{m}_\rho$ the real part of 
the complex $\rho$ pole position), obtained from the 
BaBar data set for the various models used.
The one-sigma contours for $G$ and $\tilde{T}$ are shown 
in Fig.~\ref{Babar09MOW}. 
The corresponding contours for the CMD-2(98) and SND data sets 
are shown for comparison in Fig.~\ref{CMD2SNDMOW}. 
Readers are directed to Ref.~\cite{wm10} for full details.  

\begin{table}[h]
\caption{\label{table3}
Orsay phase and separated mixing and direct $\omega\pi\pi$
coupling parameters for the BaBar(09) data. }
\vskip .1in
\begin{tabular}{|c||c|c|c|c|c|}
\hline
Parameter & KS & HLS & GS & GP/CEN$^+$ & GP/CEN$^{++}$ \\
\hline
$\phi$ (deg) & $108 \pm 1$ & $108 \pm 1$ & $107 \pm 1$ & $107 \pm 1$ & $107 \pm 1$ \\
\hline
$G$ & $0.028\pm 0.013$ & $0.035\pm 0.013$ & $0.036\pm 0.013$ & $0.039\pm 0.015$ & $0
.040\pm 0.013$ \\
$\tilde{T}$ & $-0.037\pm 0.002$ & $-0.038\pm 0.002$ & $-0.038\pm 0.002$ & $-0.0381\pm
 0.0009$ & $-0.038\pm 0.001$ \\
\hline
\end{tabular}
\end{table}

The results of Table~\ref{table3} should be compared to
those in Tables VI to IX of Ref.~\cite{wm10}.  It is immediately
apparent that the lower statistical uncertainty of the BaBar data translates
into much greater precision in the extracted value of $\tilde{T}$.  Two further
significant differences concern the central values of $\phi$ and $G$, which 
are both lower for the BaBar data compared to the other data sets.  
The BaBar data also significantly improves the significance of the evidence
for $G\neq 0$.  In Ref.~\cite{wm10} we presented combined
averages both including the KLOE data and excluding it.  
The high precision BaBar data now so dominates the 
combined averages that there is little distinction between the 
results obtained including or excluding KLOE;
we thus present only the former
in Table~\ref{table4} below.

\begin{table}[h]
\caption{\label{table4}
Combined averages including/excluding BaBar data.}
\vskip .1in
\begin{tabular}{|c|c|c|}
\hline
 & This work & Ref.~\cite{wm10} \\
\hline
\multicolumn{3}{|c|}{KLOE Included}\\
\hline
$\phi$ & $109.0^\circ \pm 1.9^\circ_{model} \pm 0.8^\circ_{data}$ & 
$113^\circ \pm 4^\circ_{model} \pm 2^\circ_{data}$ \\
\hline
$\tilde{T}$ & $-0.041 \pm 0.003_{model} \pm 0.001_{data}$ & 
$-0.044 \pm 0.006_{model} \pm 0.002_{data}$ \\
\hline
$G$ & $0.054\pm 0.014_{model} \pm 0.010_{data}$ & 
$G = 0.080\pm 0.026_{model} \pm 0.015_{data}$ \\
\hline
\end{tabular}

\end{table}

Note that the lower model dependence shown in the first column of 
Table~\ref{table4} reflects the dominance of the high-precision BaBar data
over the other data sets in the averages, rather than any improved model
consistency.  
The combined average for the complex-valued off-diagonal part of the physical
$\rho-\omega$ self-energy matrix, $\Pi_{\rho\omega}(m_\rho^2)$, now including 
the BaBar data, is
\begin{equation}
\Pi_{\rho\omega}(m_{\rho}^2) =
(-4620\pm 220_{\rm model}\pm 170_{\rm data}) + (-6100\pm 1800_{\rm model}\pm 1110_{\rm data})i\;{\rm MeV}^2 .
\label{selfenergy}
\end{equation}

In conclusion, we have updated the determination of 
$\left[\delta a_{\mu}\right]^{LO}_{had;mix}$ and the separation of $\rho-\omega$
interference in the $e^+e^-\to\pi^+\pi^-$ cross-sections into direct and mixing
induced terms using the recently released BaBar ISR data.  The main results are 
given in Eqs.~\ref{CVall} and \ref{selfenergy}, and in Table~\ref{table3}.  We conclude
that, while not at present dominant, the model-dependence of 
$\left[\delta a_{\mu}\right]^{LO}_{had;mix}$ given in Eq.~\ref{CVall}
will eventually represent a 
fundamental limitation on the use of $\tau$ data in the evaluation of $a_\mu$.

\begin{acknowledgments}
KM would like to acknowledge the hospitality of the CSSM, University of
Adelaide, and the ongoing support of the Natural Sciences and Engineering
Research Council of Canada.
\end{acknowledgments}


\end{document}